\documentclass[onecolumn,pre,notitlepage,aps,floatfix,superscriptaddress,longbibliography]{revtex4-1}
\pdfoutput=1 
\usepackage{amsmath,amssymb,eucal,graphicx,float,epstopdf,xparse}
\usepackage[colorlinks=true, urlcolor=blue, anchorcolor=blue, citecolor=blue,filecolor=blue,linkcolor=blue,menucolor=blue]{hyperref}

\usepackage{graphicx}
\usepackage{dcolumn}
\usepackage{amssymb}
\usepackage{bm}
\usepackage{color}
\usepackage{enumitem}
\usepackage{mathtools}
\usepackage{comment}
\usepackage{subfigure}
\usepackage{float}

\begin{document}

\title{The Magic of Networks Grown by Redirection}

\author{P. L. Krapivsky}
\affiliation{Department of Physics, Boston University, Boston, MA 02215, USA}
\affiliation{Santa Fe Institute, 1399 Hyde Park Road, Santa Fe, NM 87501, USA}
\author{S. Redner}
\affiliation{Santa Fe Institute, 1399 Hyde Park Road, Santa Fe, NM 87501, USA}

\begin{abstract}
  We highlight intriguing features of complex networks that are grown
  by redirection.  In this mechanism, a target node is chosen
  uniformly at random from the pre-existing network nodes and the new
  node attaches either to this initial target or to a neighbor of this
  target.  This exceedingly simple algorithm generates preferential
  attachment networks in an algorithmic time that is linear in the
  number of network nodes $N$.  Even though preferential attachment
  ostensibly requires global knowledge of the network,
  redirection requires only local knowledge.  We also show that
  changing just a single attachment rate in linear preferential
  attachment leads to a non-universal degree distribution.  Finally,
  we present unexpected consequences of redirection in networks with
  undirected links, where highly modular and non-sparse networks
  arise.
\end{abstract}

\maketitle

\section{Introduction}
\label{sec:intro}

\begin{figure}[ht]
\centerline{\includegraphics[width=0.3\textwidth]{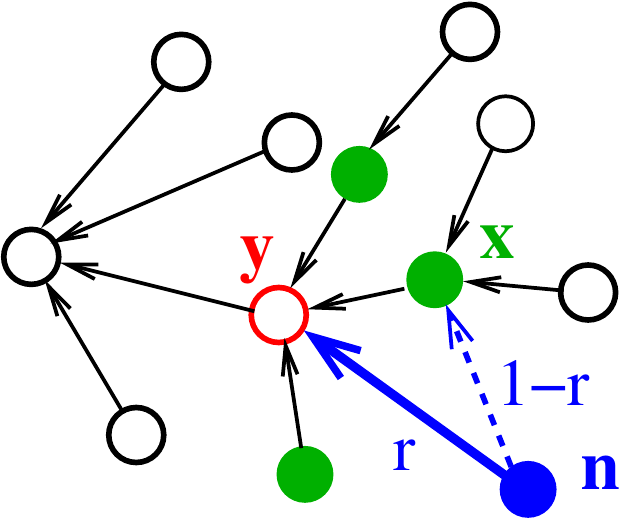}}
\caption{\small Illustration of redirection.  The rate at which a new
  node $\mathbf{n}$ attaches to the ancestor $\mathbf{y}$ of node
      $\mathbf{x}$ by redirection is proportional to the number of
      upstream neighbors of $\mathbf{y}$ (green nodes).
      \label{fig:redir}}
\end{figure}

Redirection is a natural mechanism to create growing networks.  In a
social setting, you may meet somebody and ultimately befriend of one
the friends of your initial acquaintance.  This redirection also
underlies a growth mechanism in Facebook, where you are encouraged to
create new links to some of the friends of your initial Facebook
friend~\cite{10.1257/aer.97.3.890,TRAUD20124165}.  The simplest
implementation of redirection for networks where each link has a
prescribed directionality is the following (Fig.~\ref{fig:redir}):
\begin{enumerate}
\itemsep -0.5ex
\item A new node {\tt n} picks a pre-existing node {\tt x} from the
  network uniformly at random.
\item With probability $0<1-r<1$, {\tt n} attaches to ${\tt x}$.
\item Otherwise, with probability $r$, {\tt n} attaches to the (unique) 
    ancestor node ${\tt y}$ of ${\tt x}$.
\end{enumerate}
These steps are repeated until a network of a desired size is
generated.

By construction, a network with a tree topology always remains a tree.
While it is straightforward to generalize to networks with loops by
the new node choosing multiple provisional targets, we focus on trees
both for their simplicity and because they illustrate many of the
intriguing features of networks that are grown by preferential
attachment.

Without the redirection step, the above growth rules define a model
that is known as the random recursive tree (RRT).  We discuss this
fundamental null model~\cite{Otter} in Sec.~\ref{sec:RRT}.
Redirection represents a minimalist extension of the RRT; this idea
was suggested in~\cite{Kleinberg} and developed
in~\cite{krapivsky2001organization}.  (Alternative extensions of
growth mechanisms that are still local in character
\cite{PhysRevE.67.056104,aldridge2005scale,ELMR15,PhysRevE.90.042806}
have also yield networks with broad degree distributions.)  As we well
discuss, standard redirection is equivalent to shifted linear
preferential attachment~\cite{krapivsky2001organization}, in which the
rate of attaching to a pre-existing network node of degree $k$ is
proportional to $k+\lambda$, with $\lambda =\frac{1}{r}-2$.  This
connection highlights an fascinating aspect of redirection---it
transforms a purely local growth mechanism---namely the RRT plus
redirection to the ancestor---into the global mechanism of linear
preferential attachment.  The motivation for preferential attachment
stems from the ``rich get richer''
parable~\cite{yule1924,simon1955class}; that is, popular high-degree
nodes are more likely to attract additional links merely by virtue of
being popular.  While enormous effort has been devoted to
understanding the properties of these types of networks (see, e.g.,
Refs.~\cite{barabasi99,krapivsky2000connectivity,dorogovtsev2000structure,albert2002statistical,dorogovtsev2002evolution,boccaletti2006complex,barrat2008dynamical,Newman-book}),
we will present, in Sec.~\ref{sec:PA}, a number of surprising and
under-appreciated features of preferential attachment.

As we will discuss in this section, networks that are built by the
redirection mechanism of Fig.~\ref{fig:redir} have a degree
distribution that possesses a non-universal algebraic tail,
\begin{align}
\label{Nk-PA}
N_k\sim \frac{N}{k^{\nu}}\,, \qquad \nu = 1+\frac{1}{r}>2\,.
\end{align}
The exponent must satisfy $\nu>2$ for all sparse networks whose degree
distribution has an algebraic tail. This bound follows from the
identity $\sum_{k\geq 1}kN_k=2L$ and the linear growth of the number
of links $L$ with $N$, which is the defining property of sparse
networks.  For trees, in particular, $\sum_{k\geq 1}kN_k=2L=2(N-1)$.

However, a disconcerting feature of several complex
networks~\cite{Blattner13} is that they are apparently characterized
by degree distributions with tail exponent $\nu<2$, which violates the
bound in Eq.~\eqref{Nk-PA}.  Mathematically, this implies that the sum
$\sum_{k\geq 1}kN_k$ grows superlinearly with $N$, which cannot occur
in sparse networks with $N_k\sim N/k^\nu$.  An exponent value $\nu<2$
may arise in densifying networks~\cite{lambiotte2016structural,
  bhat2016densification,Kristina}, where $L$ increases superlinearly with
$N$. Intriguingly, such an anomalously small exponent also occurs in
undirected growing trees that are generated by complete
redirection (Sec.~\ref{sec:redir-iso}).  To be consistent with the
constraint $\sum_{k\leq N}kN_k\sim N$, the amplitude of the degree
distribution must grow sublinearly with $N$, namely
\begin{equation}
\label{Nk:large}
N_k\sim \frac{N^{\nu-1}}{k^{\nu}}\,,~\qquad \nu <2\,.
\end{equation} 

Networks grown by this parameter-free complete redirection mechanism:
(a) are highly modular; (b) have numerous macrohubs; (c) consist
almost entirely of leaves (nodes of degree 1); (d) the ``core'' of the
network (nodes of degree $k\geq 2$) comprises a vanishingly small
fraction of the network as $N\to\infty$; and (e) are
non-self-averaging, namely, basic characteristics, such as $N_k$ for
any $k>1$, exhibit huge fluctuations from realization to realization.
In spite of the simplicity of complete redirection, there is little
analytical understanding of its intriguing consequences and these
represent an appealing future challenge.

We emphasize that the redirection algorithm is extremely efficient.
To build a network of $N$ nodes requires a computation time that
scales linearly with $N$, with a prefactor of the order of
one. Redirection also allows one to build networks with more general
preferential attachment mechanisms, such as sublinear preferential
attachment, with nearly the same efficiency as the original
redirection algorithm (Sec.~\ref{sec:redir-deg}).

\section{The Random Recursive Tree (RRT)}
\label{sec:RRT}

We begin our discussion with the RRT, first introduced by
Otter~\cite{Otter}, in which nodes are added to the network one by
one. Each new node attaches to a single ``target'' node that is chosen
uniformly at random among the already existing nodes; that is, the
attachment rate $A_k=1$, for any degree $k$.  By the restriction that each
new node has a single attachment point (equivalently, the out degree
of every node equals 1), the resulting network is a tree.  If a new
node attaches to more than one pre-existing node, loops could form.
The degree distribution of a network with loops is modified only in
the amplitude of the degree distribution compared to growing trees.
On the other hand, topological features of networks with loops are
different than trees, but our focus is on the degree distribution, for
which it is simplest to focus on tree networks.

The growth rules of the RRT thus are:
\begin{enumerate}
\itemsep -0.5ex
\item Pick one of the nodes of the RRT---defined as the target---with uniform probability.
\item Introduce a new node that links to the target node.
\end{enumerate}
Starting with a single node, these two steps are repeated until the
tree reaches a desired number of nodes $N$.

\begin{figure}[ht]
  \begin{center}
    \includegraphics[width=0.2\textwidth]{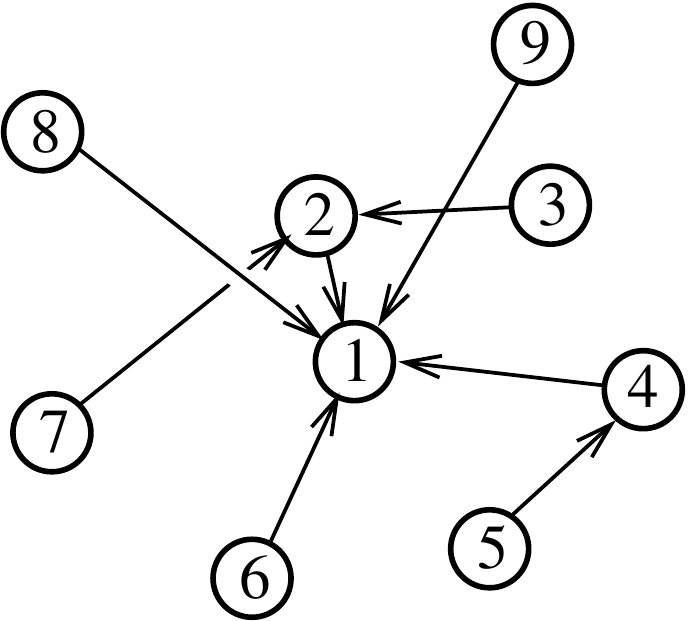}
    \caption{\small A random recursive tree of 9 nodes, showing the
      ordering of the nodes and each of their attachment points.}
\label{RRT}
  \end{center}
\end{figure}

\subsection*{The degree distribution}
\index{Networks!degree distribution}

We first outline how to derive the exact degree distribution and then
determine the degree distribution in the limit $N\to\infty$.  The
degree state of any network is characterized by the vector
${\bf N}\equiv \{N_1,N_2,\ldots\}$, where $N_k$ denotes the number of
nodes of degree $k$.  When a new node is introduced, the changes in
the network state vector ${\bf N}$
are~\cite{KR02-fluct,KRB}:
\begin{equation}               
\label{N12}
\begin{split}
&\mathrm{attach\ to\ node\ of\ degree\ 1\!:}~~~~~~~~\, (N_1,N_2)\to (N_1,N_2+1)\\
&\mathrm{attach\ to\ node\ of\ degree\ } k>1\!:~~ (N_1,N_k,N_{k+1})\to  (N_1+1,N_k-1,N_{k+1}+1)\,,
\end{split}
\end{equation}
while the state of all other network nodes are unchanged.  Typically
we are not interested in the full probability distribution
$P({\bf N})$, but just the average number of nodes of a given degree,
$\langle N_k\rangle$, namely, the degree distribution; the angle
brackets denote an average over all possible growth histories of the
network.

Let us determine how the $N_k$ change when a new node is added to the
network.  As indicated by Eq.~\eqref{N12}, we need to separately
consider nodes of degree 1 and nodes of degree greater than 1.  The
number of nodes of degree 1, $N_1(N)$, i.e., the number of leaves, is a random variable that
changes with each node addition according to
\begin{equation}
\label{N1}
N_1(N\!+\!1)=\begin{cases}
N_1(N)   & \qquad \mathrm{probability}\quad \frac{N_1}{N}\\[2mm]
N_1(N) +1 & \qquad\mathrm{probability}\quad 1-\frac{N_1}{N}\,.
\end{cases}                                   
\end{equation}
These equations apply for $N\geq 2$, while the natural initial
condition is $N_1(2)=2$.  This equation expresses the two
possibilities when a new node joins the network: with probability
$N_1/N$, the new node attaches to a node of degree 1 and the number of
such nodes does not change, while with probability $(1-N_1/N)$, the
new node attaches to a node of degree $k>1$ and $N_1$ increases by 1.

The evolution equation for the average number of leaves is therefore 
\begin{align}
\label{N1av}
\big\langle N_1(N\!+\!1)\big\rangle
&=\Big\langle N_1(N)\times\frac{N_1(N)}{N}\Big\rangle
+\Big\langle \left(N_1(N)+1\right)\times \Big(1-\frac{N_1(N)}{N}\Big)\Big\rangle\nonumber \\
&=1+\Big(1-\frac{1}{N}\Big)\big\langle N_1(N)\big\rangle\,.
\end{align}
Because the relevant time-like variable that characterizes the network
size is the total number of nodes $N$, we will always use $N$ as the
time variable.  The solution to this recursion, for $N\geq 2$, is
\begin{equation}
\label{N1av-sol}
\big\langle N_1(N)\big\rangle = \frac{N}{2}+\frac{1}{N-1}~.
\end{equation}

The discrete approach can be used to determine higher moments of the
random variable $N_1(N)$.  The second moment $\langle N_1^2(N)\rangle$
is especially important as we can obtain the variance and
thereby quantify degree fluctuations.  From Eq.~\eqref{N1}, we deduce the recurrence
for the second moment 
\begin{align*}
\left\langle N_1^2(N+1)\right\rangle
&=1+\left(1-\frac{2}{N}\right)\langle N_1^2(N)\rangle
+\left(2-\frac{1}{N}\right)\langle N_1(N)\rangle\,,
\end{align*}
whose solution is
\begin{align}
\label{N2av}
  \left\langle N_1^2(N)\right\rangle = \frac{N(3N+1)}{12}+ \frac{N}{N-1}~.
\end{align}
From the first two moments, the variance, for $N\geq 3$, is
\begin{align}
\label{N1var-sol}
\left\langle N_1^2(N)\right\rangle_c \equiv \left\langle N_1^2(N)\right\rangle- \left\langle
  N_1(N)\right\rangle^2=\frac{N}{12}- \frac{1}{(N-1)^2}~,
\end{align}
so the deviation of $N_1(N)$ from its average is of the order of $\sqrt{N}$. 
Higher cumulants of the number of leaves also grow as $\sqrt{N}$. 
The cumulants $\left\langle N_1^p(N)\right\rangle_c$ 
with arbitrary integer $p\geq 1$ are given by remarkably simple formula
\begin{equation}
\label{LN-p}
\left\langle N_1^p(N)\right\rangle_c  = p^{-1} B_p  N + \frac{(-1)^{p-1} (p-1)! }{(N-1)^p}
\end{equation}
applicable when $N\geq p+1$. 
Here $B_p$ are Bernoulli numbers defined \cite{Knuth} as the coefficients in the power series 
\begin{equation*}
\label{B:def}
\frac{z}{e^z-1}+z =\sum_{p\geq 0}B_p\,\frac{z^p}{p!}
\end{equation*}

Thus for large $N$, the number of nodes of degree 1 is sharply distributed about its
average value. For this reason, one may ignore fluctuations and focus
on the average.  This same holds for all nodes of higher degrees as long as the 
number of such nodes is large, $N_k(N)\gg 1$. Thus we again focus on the average.

By similar reasoning as that used for $N_1$, the number of nodes of
degree $k\geq 2$, evolves according to
\begin{equation}                                                                                
\label{Nk-cases}
N_k(N\!+\!1)=
\begin{cases}
  N_k(N)-1   & \qquad\mathrm{probability}\quad \frac{N_k}{N}\\[2mm]
  N_k(N)+1   & \qquad\mathrm{probability}\quad \frac{N_{k-1}}{N}\\[2mm]
  N_k(N) & \qquad\mathrm{probability}\quad 1-\frac{N_{k-1}+N_k}{N}
\end{cases}                                       
\end{equation}
after each node addition.  Following the same steps that led to
Eq.~\eqref{N1av}, the evolution equation for  $\langle N_k\rangle$ is
\begin{equation}
\label{Nkav}
\left\langle N_k(N\!+\!1)\right\rangle
=\left\langle N_k(N)\right\rangle 
+\bigg\langle \frac{N_{k-1}(N)-N_k(N)}{N}\bigg\rangle.
\end{equation}
While this equation can again be solved to give the exact degree
distribution for finite networks, we now restrict ourselves to the
leading behavior of the degree distribution for $N\to\infty$.  For
simplicity, we drop the angle brackets and the argument $N$, so that
we write $N_k$ for the average number of nodes of degree $k$ in a
network that contains $N$ nodes.  Next, we replace the discrete
differences with derivatives in Eqs.~\eqref{N1av} and \eqref{Nkav}, so
that the asymptotic degree distribution evolves according to the
master equation
\begin{subequations}
\begin{equation}
\label{ugt-Nk}
\dot N_k\equiv \frac{d N_k}{dN}= \frac{N_{k-1}- N_k}{N}+\delta_{k,1}~.
\end{equation}
The first equation is $\dot N_1 =-{N_1}/{N}+ 1$, with solution $N_1= N/2$.  Then  $\dot N_2=(N_1- N_2)/{N}$, with solution
$N_2= N/4$.  Continuing one finds that all the $N_k$ are proportional
to $N$.  Thus we write $n_k\equiv N_k/N$ and reduce Eq.~\eqref{ugt-Nk} to
\begin{equation}
\label{ugt-nk}
 n_k= n_{k-1}- n_k+\delta_{k,1}
\end{equation}
\end{subequations}
leading to the exponential degree distribution $n_k= 2^{-k}$.

\section{Preferential Attachment}
\label{sec:PA}

In preferential attachment, the rate $A_k$ at which a node attaches to
a pre-existing node of degree $k$ is an increasing function of $k$.  A
ubiquitous feature of preferential attachment networks is that their
degree distributions have broad tails, a fact that sparked much
interest in this class of networks over the past two decades.  We now
derive this scale-free degree distribution using the approach of in
Ref.~\cite{krapivsky2001organization}.

\subsection*{Master equation}
The evolution of the degree distribution for a network whose growth is
governed by an attachment rate $A_k$ is (compare with Eq.~\eqref{ugt-Nk}
for the RRT):
\begin{equation}
\label{Nk}
\dot N_k= \frac{A_{k-1} N_{k-1}-A_k N_k}{A}+\delta_{k1}.
\end{equation}
The first term on the right accounts for the new node connecting to a
pre-existing node that already has $k-1$ links, thereby increasing
$N_k$ by one.  Since there are $N_{k-1}$ nodes of degree $k-1$, the
rate at which such processes occur is proportional to
$A_{k-1}N_{k-1}$.  The total rate
$A\equiv A(N)\equiv\sum_{j\geq 1} A_jN_j$ in the denominator means
that $A_{k-1}/A$ is the probability for a node of degree $k-1$ to
become a node of degree $k$.  A corresponding role is played by the
second term on the right.  The overall amplitude of $A_k$ is
immaterial, since only the ratio $A_k/A$ appears in the master
equation.  The last term accounts for the introduction of a new node
that has one outgoing link and no incoming links.  To determine the
degree distribution, we need to specify the attachment rate $A_k$.  We
focus on power-law preferential attachment, $A_k= k^\gamma$, with
$\gamma \geq 0$.  We will show that different behaviors arise for
sublinear ($\gamma<1$), superlinear ($\gamma>1$), and linear
($\gamma=1$) attachment rates.  The linear case is especially rich
because the degree distribution is nonuniversal.

When confronted with determining a non-trivial distribution, it is
often instructive to first deal with the simpler problem of
determining low-order moments of the degree distribution
$M_\alpha(N) \equiv \sum_j j^\alpha N_j$.  The zeroth and first
moments of this distribution have particularly simple $N$ dependences:
$\dot M_0= \sum_j \dot N_j =1$ and
$\dot M_1 = \sum_j j\, \dot N_j =2$.  The equation for $M_0$ states
that the total number of nodes (of any degree) increases by 1 each
time a new node is introduced.  Similarly, the equation for $M_1$
states the total degree of the network, $\sum jN_j$, increases by two
when the single link associated with the new node is added to the
network.  Since both the zeroth and first moments of the degree
distribution increase linearly with $N$, the total rate
$A=\sum_j j^\gamma N_j$ also grows linearly with $N$, because $A$ is
intermediate to the zeroth and first moments for
$0\leq \gamma \leq 1$.  Asymptotically, $A\simeq \mu N$, with the as
yet-undetermined amplitude $\mu$ that must range between 1 and 2 as
$\gamma$ increases from 0 to 1.

Solving for the first few $N_k$ from Eq.~\eqref{Nk}, it becomes clear
that each $N_k$ is also proportional to $N$.  This fact suggests
substituting $N_k(N)=n_kN$ and $A\simeq \mu N$ into these master
equations.  With this step, the overall $N$ dependence cancels,
leaving behind the recursion relations
$n_k=(A_{k-1}n_{k-1}- A_{k}n_k)/\mu$ for $k>1$ and $n_1=1-A_1n_1/\mu$.
After straightforward algebra, the degree distribution is
\begin{subequations}
\begin{equation}
  \label{Nkgen}
n_k=\frac{\mu}{A_k}\prod_{1\leq j\leq k} \left(1+\frac{\mu}{A_j}\right)^{-1}.
\end{equation}
Using the definition $\mu=\sum_{j\geq 1}A_j n_j$ in \eqref{Nkgen} we obtain
\begin{equation}
\label{mugen}
\sum_{k\geq 1} \prod_{1\leq j\leq k}
\left(1+\frac{\mu}{A_j}\right)^{-1}=1.
\end{equation}
\end{subequations}
To extract the physical meaning of the general solution \eqref{Nkgen}
with $\mu$ implicitly determined by \eqref{mugen} we examine the
asymptotic behavior for the three generic cases of sublinear,
superlinear, and linear preferential attachment.

\subsubsection{Sublinear preferential attachment}

For $A_k=k^\gamma$ with $\gamma<1$, we rewrite the product in
Eq.~\eqref{Nkgen} as the exponential of a sum of logarithms, convert
the sum to an integral, and then expand the logarithm inside the
integral in a Taylor series.  These straightforward steps lead to
\begin{eqnarray}
\label{cases} 
n_k\sim
\begin{cases}
k^{-\gamma}\exp
\left[-\mu\left(\frac{k^{1-\gamma}-2^{1-\gamma}}{1-\gamma}\right)\right]
&\qquad\frac{1}{2}<\gamma<1,\\ \\
k^{(\mu^2-1)/2}\exp\left[-2\mu\,\sqrt{k} \right] 
&~~~~~~\qquad \gamma=\frac{1}{2},\\ \\
k^{-\gamma}\exp\left[-\mu\, \frac{k^{1-\gamma}}{1-\gamma}
+\frac{\mu^2}{2}\, \frac{k^{1-2\gamma}}{1-2\gamma}\right]
&\qquad\frac{1}{3}<\gamma<\frac{1}{2}\,,
\end{cases}
\end{eqnarray}
with similar, but more complicated expressions for $n_k$ for still
smaller values of $\gamma$. Each time $\gamma$ decreases through
$\frac{1}{m}$, where $m$ is an integer, an additional term is
generated in the exponential that is an increasing function of $k$.
Nevertheless, for any value of $\gamma<1$, the leading behavior is
always the universal stretched exponential decay,
$\exp(-{\rm const.}\times k^{1-\gamma})$.

\subsubsection{Superlinear preferential attachment}

\begin{figure}[ht]
  \begin{center}
    \includegraphics[width=0.3\textwidth]{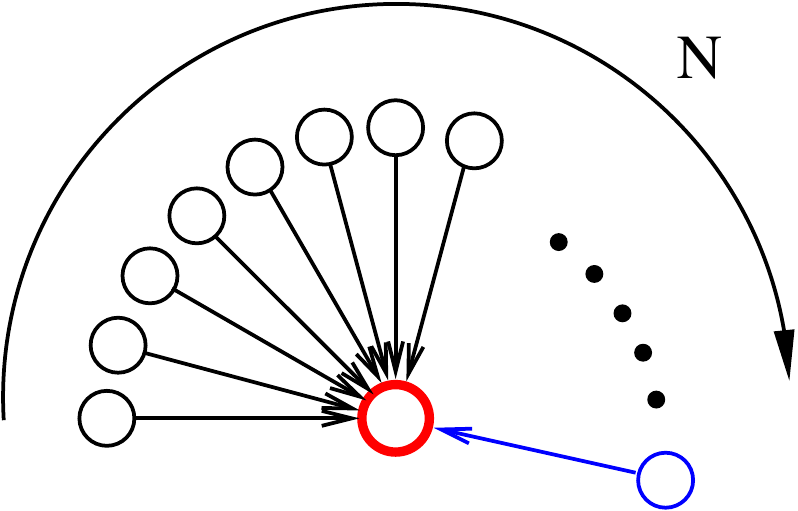}
    \caption{\small Creation of a ``bible'' in which each new node attaches only to
      the bible (red).}
    \label{fig:bible}
  \end{center}
\end{figure}

For $\gamma>1$, a gelation-like phenomenon occurs in which nearly all
links attach to a single node.  Let us first treat the ultra singular
behavior that arises for $\gamma>2$, for which there is a non-zero
probability for a ``bible'' to occur---a node that links to every
  other node in an infinite network, while only a finite number of
links exist between all other nodes.  To determine the probability for
a bible, suppose that a network of $N+1$ nodes contains a bible
(Fig.~\ref{fig:bible}).  The probability that the next node links to
the bible is $N^\gamma/(N+N^\gamma)$, and the probability that this
pattern of connections continues indefinitely is
$\mathcal{P} = \prod_{N\geq 1}N^\gamma/(N+N^\gamma)$.  Using the same
asymptotic analysis as above, where we write the product as the
exponential of a sum of logarithms, expand the logarithm for large
$N$, and approximate the sum as an integral, the asymptotic behavior of
this product is $\mathcal{P}=0$ for $\gamma\leq 2$, and
$\mathcal{P}>0$ for $\gamma>2$.  Strikingly, there a non-zero
probability for a bible to exist in an infinite network for
$\gamma>2$!  

\begin{figure}[ht]
  \begin{center}
    \includegraphics[width=0.8\textwidth]{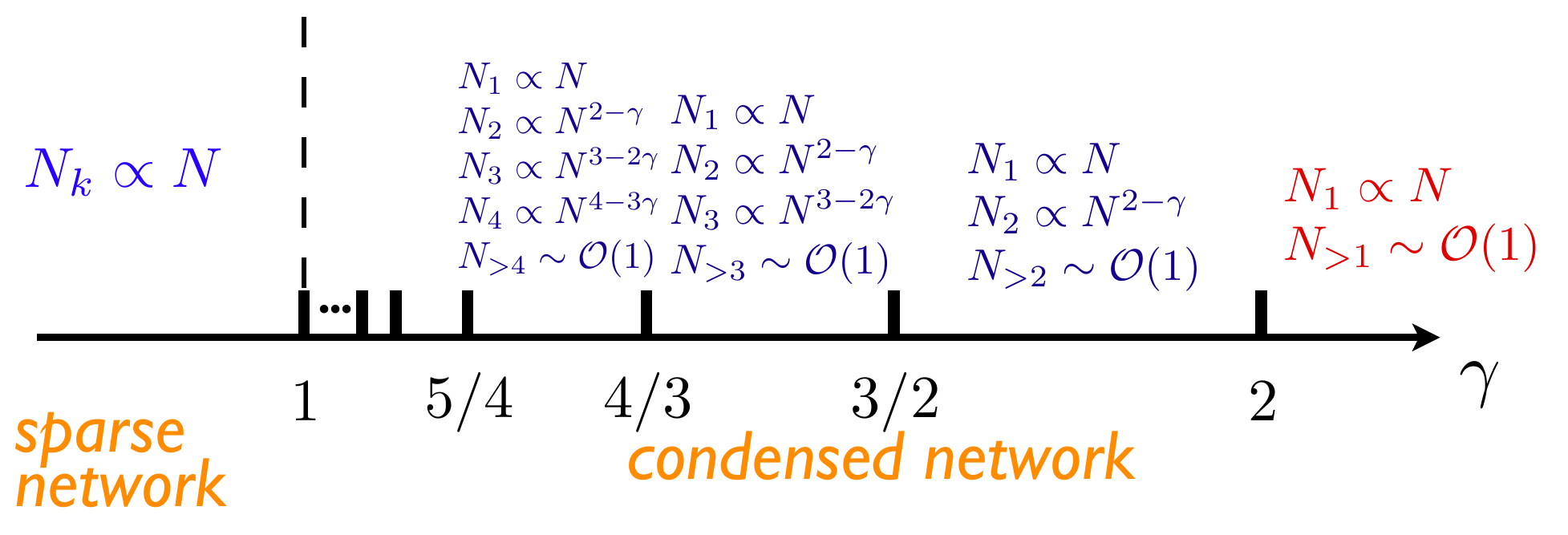}
    \caption{\small Illustration of the sequence of phase transitions
      that arise in superlinear preferential attachment.  Starting
      with an ultra-condensed network for $\gamma>2$, the network
      contains progressively more low-degree nodes each time $\gamma$
      passes through $m/(m-1)$.  The network becomes sparse when
      $\gamma$ reaches 1, where the number of nodes of any degree are
      all proportional to $N$.}
    \label{fig:superlinear}
  \end{center}
\end{figure}

When $1<\gamma<2$, the attachment pattern of low-degree nodes is not
as simple as in Fig.~\ref{fig:bible}, but there continues to be a
single node whose degree is of the order of $N$.  There is also an
infinite sequence of transition points when $\gamma$ passes through
$\frac{m}{m-1}$, with $m$ an integer greater than 2, in which number
of nodes of degree $k\leq m$ grows as $N^{k-(k-1)\gamma}$, while the
number of nodes of degree $k >m$ remain finite for $N\to\infty$
(Fig.~\ref{fig:superlinear}).  To understand this behavior in a simple
way, it is instructive to study the governing equations for each $N_k$
one by one.  For $N_1$ we have
\begin{align*}
  \dot N_1 = 1- \frac{N_1}{A}\,.
\end{align*}

We now make the assumption that the total attachment rate is governed
by the single highest-degree node, with degree of the order of $N$.
Thus $A=\sum j^\gamma N_j \sim N^\gamma$.  Since $N_1$ can at most be
of the order of $N$, the second term in the above equation is
negligible, so that $\dot N_1\sim 1$ or $N_1\sim N$. Similarly, the
equation for $N_2$ is
\begin{align*}
  \dot N_2 \simeq \frac{N_1-2^\gamma N_2}{N^\gamma}\,.
\end{align*}
Again, neglecting the second term, gives
$\dot N_2\simeq N^{1-\gamma}$, from which $N_2\sim N^{2-\gamma}$.  We
can then verify that the term that we dropped is indeed negligible.
Continuing this self-consistent procedure for general degree $k$, we
find
\begin{align}
  N_k\sim N^{k-(k-1)\gamma}\,,
\end{align}
as long as the exponent of $N_k$ is positive, while $N_k$ will be
finite for $N\to\infty$ for values of $k$ for which $k-(k-1)\gamma$ is
negative (Fig.~\ref{fig:superlinear}).

Thus we predict an infinite sequence of transitions at
$\gamma=\gamma_m =\frac{m}{m-1}$.  For $\gamma>\gamma_m$, the number
of nodes of degree $k>m$ are all of $\mathcal{O}(1)$, while nodes of
degrees $k\leq m$ grows sublinearly with $N$, as $N^{k-(k-1)\gamma}$.
This set of transitions becomes progressively more dense as
$\gamma\to 1$ from above.  At $\gamma=1$, the network changes its
character from condensed, where a hub node has degree of
$\mathcal{O}(N)$, to sparse, where the number of nodes of any degree
is proportional to $N$.

\subsubsection{Linear preferential attachment}

Here, it is important to distinguish between strictly linear
preferential attachment, $A_k=k$, and asymptotically linear
preferential attachment, $A_k\simeq k$.  In the former case, the total
attachment rate is $A=\sum_k A_k N_k=\sum_k k N_k=2N$.  Substituting
this value of $\mu=2$ into Eq.~\eqref{Nkgen} and performing some
simple algebra immediately leads to the discrete power-law form of the
degree distribution
\begin{equation}
\label{nk1}
n_k=\frac{4}{k(k+1)(k+2)}= \frac{4\,\Gamma(k)}{\Gamma(k+3)}~,
\end{equation}
where $\Gamma$ is the Euler gamma function.  From this power-law
degree distribution, the mean degree
$\langle k\rangle =\sum_{k\geq 1}k n_k=2$, as it must, but the
mean-square degree $\langle k^2\rangle=\infty$. Thus fluctuations in
the mean degree, namely, the spread in the mean degree for different
realization of large networks of $N$ nodes, diverges for $N\to\infty$.

The surprising feature of asymptotically linear preferential
attachment growth is that the degree distribution exponent is non-universal.  This non-universality is at odds with the common
wisdom of statistical physics in which the absence of a characteristic
scale leads to universal scaling properties.  One natural form for an
asymptotically linear attachment rate is $A_k= k+\lambda$, with
$\lambda$ a constant.  This modification implies that the amplitude
$\mu$ in $A=\mu N$ is no longer equal to 2, but assume a wide range of
values (see below). Now Eq.~\eqref{Nkgen} becomes
\begin{align}
\label{nk-asymp-lin}
n_k=\frac{\mu}{A_k}\prod_{1\leq j\leq k} \left(1+\frac{\mu}{A_j}\right)^{-1} 
&\sim  \frac{\mu}{k} \exp\left[-\int_1^k \ln\left(1+\frac{\mu}{j}\right)\,dj\right]\nonumber\\
&\sim \frac{\mu}{k} \exp\left[-\mu \int_1^k \frac{dj}{j}\right]\nonumber\\[1mm]
&\sim k^{-(1+\mu)}~.
\end{align}
Thus the degree exponent $\nu=1+\mu$ can take any value larger
than 2 merely by tuning the amplitude $\mu$.

As an explicit and surprising example, consider the attachment rate
$A_k=k$ for $k\geq 2$, while $A_1\equiv \alpha$ is arbitrary.  It is
now convenient to separate $A_1$ and $A_k$ for $k\geq 2$ in
Eq.~\eqref{Nkgen} to recast this equation as
  \begin{equation}
\label{mu2}
\mu=A_1\sum_{k=2}^\infty \prod_{j=2}^{k}
\left(1+\frac{\mu}{A_j}\right)^{-1}
=\;\alpha\;\sum_{k=2}^\infty \Gamma(2+\mu)\,
\frac{\Gamma(1+k)}{\Gamma(1+\mu+k)},
\end{equation}
where we express the product as the ratio of gamma functions.

The sum can be evaluated by employing the
identity \cite{Knuth}
\begin{equation*}
\sum_{k=2}^\infty \frac{\Gamma(a+k)}{\Gamma(b+k)}=
\frac{\Gamma(a+2)}{(b-a-1)\Gamma(b+1)}~,
\end{equation*}
so that Eq.~\eqref{mu2} becomes $\mu(\mu-1)=2\alpha$, with solution
$\mu=(1+\sqrt{1+8\alpha})/2$.  Thus the degree exponent $\nu=1+\mu$ is
\begin{equation}
\label{nu}
\nu=\frac{3+\sqrt{1+8\alpha}}{2}~.
\end{equation}
As examples, the degree distribution exponent is $\nu=4$ for
$\alpha=3$ and $\nu=5/2$ for $\alpha=3/8$.  For $0<\alpha<1$, the
exponent lies in the range $2<\nu<3$, while for $\alpha>1$, $\nu>3$.
While the degree distribution exponent must satisfy the lower bound
$\nu>2$, there is no upper bound for $\nu$; in particular,
$\nu\to \sqrt{2\alpha}$ as $\alpha\to \infty$.  We emphasize that
changing just a single attachment rate leads to a global
effect on the degree distribution.  This global effect arises because
the amplitude $\mu$ appears inside the infinite product in
Eq.~\eqref{Nkgen}.  This multiplicative nature strongly affects the
degree distribution itself and thereby the degree distribution
exponent.

\section{Network Growth by Redirection}
\label{sec:redir}

We now discuss a deceptively simple modification of the RRT with
profound consequences.  This is the notion of redirection where
a new node may attach to a pre-existing target node, or it to a
neighbor of this target~\cite{Kleinberg,krapivsky2001organization}.

\subsection{Constant redirection probability} 

First we treat the redirection
algorithm~\cite{krapivsky2001organization} that was outlined in the
introduction.  There is one subtlety in this algorithm because
redirection requires that every node has an ancestor.  To ensure this
condition always holds, the initial state, for example, could consist
of at least two nodes and one link, with each node defined as the
ancestor of the other. Other simple starting graphs are equally
suitable, such as a triangle with cyclic links.  

According to the redirection algorithm, the degree distribution
evolves according to
\begin{subequations}
\begin{align}
\label{NkC}
\dot N_k= \frac{1-r}{N}\Big[N_{k-1}-N_k\Big]
+\frac{r}{N}\Big[(k-2)N_{k-1}-(k-1)N_k\Big] +\delta_{k,1}\,.
\end{align}
The terms within the first square brackets correspond to attachment to
the initially selected node, whose evolution equation is just that of
the RRT (Eq.~\eqref{ugt-Nk}) for redirection probability $r=0$.  The
terms within the second square brackets account for the change in
$N_k$ due to redirection.  To understand their origin, consider first
the gain term.  Since the initial node is chosen uniformly, if
redirection does occur, then the probability that a node of degree
$k-1$ receives the newly redirected link is proportional to the number
of its upstream neighbors (green nodes in Fig.~\ref{fig:redir}), which
equals $k-2$.  A parallel argument applies for the redirection-driven
loss term.  The crucial point is that the rate at which attachment
occurs to a given node is proportional to the number of its upstream
neighbors, which, in turn, is proportional its degree.  Thus linear
preferential attachment is implicit in this purely local redirection
rule.

The redirection mechanism has an unexpected connection to the
friendship paradox~\cite{hodas2013friendship, eom2014generalized},
which states that the neighbors of a randomly selected node are more
popular (have higher degrees), on average, than the initially selected
node.  As illustrated in Fig.~\ref{fig:redir}, there are three
distinct ways to attach node $\mathbf{y}$ by redirection from upstream
nodes.  The higher the degree of node $\mathbf{y}$, the more likely
attachment to it by redirection occurs.  Thus we expect that node
$\mathbf{y}$ will have more neighbors, on average, than the initial
node $\mathbf{x}$

By a straightforward rearrangement of terms, \eqref{NkC} may be
re-expressed as
\begin{align}
\dot N_k&= \frac{r}{N}\left\{\left[k-1+\left(\frac{1}{r}-2\right)\right]N_{k-1}
                 -\left[k+\left(\frac{1}{r}-2\right)\right]N_k\right\}+\delta_{k,1}\nonumber\\[2mm]
  &\equiv  \frac{1}{A}\Big\{\left(k-1+\lambda\right)N_{k-1}
                 -\left(k+\lambda\right)N_k\Big\}+\delta_{k,1}\,,
\end{align}
with $\lambda=\frac{1}{r}-2$ and total attachment rate
$A=N/r= (2+\lambda)N$.  Thus uniform attachment, in conjunction with
redirection, generates shifted linear preferential attachment,
with $A_k = k +\lambda$.  The particular case of strictly
linear preferential attachment arises for the choice $r=\frac{1}{2}$.
When we now substitute attachment rate $A_k=k+\lambda$ and
$\mu=2+\lambda$ into the general formula \eqref{Nkgen} for the degree
distribution, we obtain
\begin{align}
  n_k=\frac{\mu}{A_k}\prod_{1\leq j\leq k} \left(1+\frac{\mu}{A_j}\right)^{-1}
     = (2+\lambda) \frac{\Gamma(3+2\lambda)}{\Gamma(1+\lambda)}
       \frac{\Gamma(k+\lambda)}{\Gamma(k+3+2\lambda)}
  \sim k^{-(3+\lambda)}\,.
\end{align}
\end{subequations}
Since the redirection probability lies between 0 and 1, the additive
shift $\lambda$ lies between $-1$ and $\infty$.  Thus the degree
distribution exponent can take on any value that is greater
than 2.  In the extreme case of $r=1$ a star-like network arises whose
detailed structure depends on the initial condition.

It is also worth mentioning the many intriguing results that emerge
from simple extensions of this redirection mechanism.  Starting with
the RRT, each node has a genealogical tree of ancestors.  It is
natural to grow a network in which redirection can occur equiprobably
to any node in the genealogical tree of an initial target
node~\cite{BK10}, or to all nodes in this genealogical
tree~\cite{KR05}.  The latter leads to a network that is no longer
sparse, as the number of links $L$ grows as $N\ln N$.  Amusingly, this
redirection mechanism to all ancestors is isomorphic to a basic
hypergraph model, known as the random recursive
hypergraph~\cite{PK23}.

Finally, we wish to emphasize the extreme simplicity of this
redirection algorithm.  Each node addition requires only two elemental
operations: (i) select a target node, and (ii) choose to attach either
to this target or to its ancestor.  This algorithm allows one to
generate a network of $N$ nodes in roughly $2N$ algorithmic steps.  It
is therefore possible to quickly generate very large networks.
Crucially, a purely local rule---tracking the ancestor of each
node---is equivalent to the global rule that underlies
preferential attachment.  Ostensibly, one needs to know the degrees of
all the nodes in the network to implement preferential attachment.  As
the redirection algorithm shows, this global information is not
needed.

\subsection{Degree-based redirection}
\label{sec:redir-deg}

\begin{figure}[ht]
\begin{center}
\includegraphics[width=0.275\textwidth]{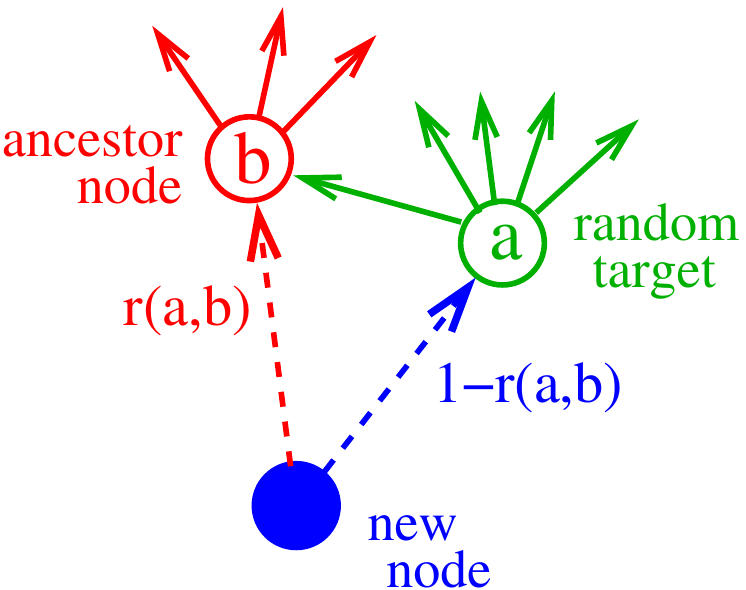} 
\caption{\small Illustration of degree-based redirection.  A new node (blue)
  attaches to a random target of degree $a$ with probability
  $1-r(a,b)$ and attaches to the ancestor node (degree $b$) of the
  target with probability $r(a,b)$.}
  \label{fig:rab}
\end{center}
\end{figure}

To illustrate the utility and generality of redirection, we exploit
the local information that is readily available---the degree $a$ of
the initial target node and the degree $b$ of the ancestor---to
efficiently generate sublinear preferential attachment networks.  In
degree-based redirection~\cite{gabel2013sublinear}, we merely define
the redirection probability $r$ to be a suitably chosen function of
these two degrees $a$ and $b$; that is $r=r(a,b)$ (see
Fig.~\ref{fig:rab}).

To show how sublinear preferential attachment can be achieved from
this still-local information, we define $f_k$ as the total probability
that an incoming link is redirected from a randomly selected
target node of degree $k$ to the parent of this target.  Similarly, we
define $t_k$ as the total probability that an incoming link is
redirected to a parent node of degree $k$ after the incoming
node initially selected one of the child nodes of this parent.
Formally, these probabilities are defined in terms of the redirection
probabilities by
\begin{equation}
\label{prob}
f_k=\sum_{b\geq 1}\frac{r(k,b)N(k,b)}{N_k}\,,\qquad \qquad 
t_k=\sum_{a\geq 1}\frac{r(a,k)N(a,k)}{(k-1)N_k}\,,
\end{equation}
where $N_k=\sum_{b\geq 1} N(k,b)$ and $N(a,b)$ is the correlation function
that specifies the number of nodes of degree $a$ that have a parent of degree
$b$.  Thus $f_k$ is the mean redirection probability averaged over all $N_k$
possible target nodes of degree $k$.  Likewise, since each node of degree $k$
has $k-1$ children, there are $(k-1)N_k$ possible target nodes whose
redirection probabilities are averaged to give $t_k$.

In terms of these probabilities $f_k$ and $t_k$, the master equation that
governs the evolution of $N_k$ is
\begin{equation}
\label{masterRA}
\dot N_k=\frac{(1\!-\!f_{k\!-\!1})N_{k-1}-(1\!-\!f_k)N_k}{N} 
+\frac{(k\!-\!2)t_{k-1}N_{k\!-\!1}-(k-1)t_{k}N_{k}}{N} +\delta_{k,1}.
\end{equation}
The first ratio corresponds to instances of network growth for which
the incoming node actually attaches to the initial target.  For
example, the term $(1-f_k)N_k/N$ gives the probability that one of the
$N_k$ target nodes of degree $k$ is randomly selected and that the
link from the new node is not redirected away from this target.
Similarly, the second ratio corresponds to instances in which the link
to the target node is redirected to the ancestor.  For example,
the term $(k-1)t_kN_k/N$ gives the probability that one of the
$(k-1)N_k$ children of nodes of degree $k$ is chosen as the target and
that the new node is redirected.  Lastly, the term
$\delta_{k,1}$ accounts for the newly added node of degree 1.

By rearranging terms, we express \eqref{masterRA} in the generic
form of Eq.~\eqref{Nk}, with the attachment rate
\begin{subequations}
\begin{equation}
\label{Ak}
\frac{A_k}{A}=\frac{(k-1)t_k+1-f_k}{N}\,.
\end{equation}
Since the quantities $f_k$ and $t_k$ are normalized probabilities, the
asymptotic behavior of the above expression is $A_k\sim k\,t_k$.  Thus
a redirection probability $r(a,b)$ for which $t_k$ is a decreasing
function of $k$ will asymptotically correspond to sublinear
preferential attachment.  A natural choice for such a redirection
probability is $r(a,b)=b^{\gamma-1}$, with $0<\gamma<1$, so that the
redirection probability decreases as the degree of the parent node
increases.  Because $r$ depends only on the degree of the parent node
(Fig.~\ref{fig:rab}), Eq.~\eqref{prob} reduces to $t_k=k^{\gamma-1}$.
Using this form of $t_k$ in Eq.~\eqref{Ak} yields
\begin{equation}
\label{SubLinear}
\frac{A_k}{A}=\frac{k^{\gamma}-k^{\gamma-1}+1-f_k}{N}\,,
\end{equation}
\end{subequations}
whose leading behavior is indeed sublinear preferential attachment,
$A_k\sim k^{\gamma}$.  This equivalence to sublinear
preferential attachment allows to generate a network of $N$ nodes with
a stretched exponential degree distribution in an algorithmic time that
is also of the order of $N$.

What happens in the opposite case of enhanced redirection, in
which the redirection probability is an increasing function of the
degree of the parent node~\cite{gabel2013sublinear,gabel2014highly}? This
attachment rule leads to highly modular networks that contains
multiple macrohubs, with most nodes having degree 1 (leaves).
Furthermore, the degree distribution exhibits the anomalous scaling
given in Eq.~\eqref{Nk:large}, with $\nu$ strictly less than
2. Similar phenomenology also occurs in the simpler example of
redirection rule for undirected networks (see below).

\subsection{Complete redirection in undirected networks}
\label{sec:redir-iso}

Link directionality is important in social and technological networks,
but there are many situations where networks are
undirected~\cite{Diestel,Drmota,Frieze}.  The influence of redirection
on undirected networks is profound and there is little analytical
understanding of this enigmatic case.

\begin{figure}[ht]
\centerline{\subfigure[]{\includegraphics[width=0.25\textwidth]{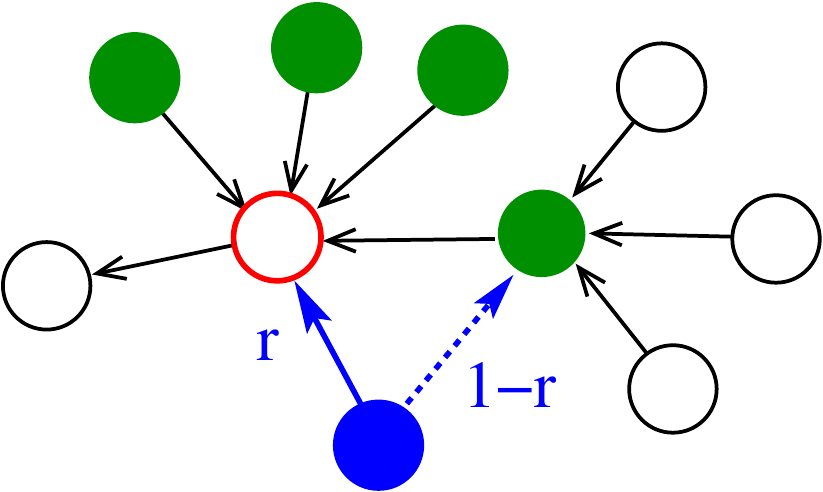}}\qquad\qquad\qquad
\subfigure[]{\includegraphics[width=0.25\textwidth]{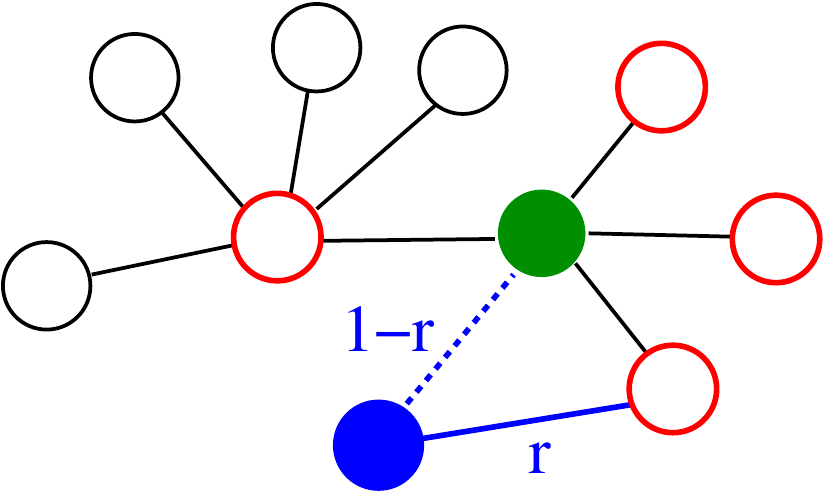}}}
\caption{\small Comparison of redirection for (a) directed and (b)
  undirected networks.  (a) The new node (blue) attaches by
  redirection to the unique ancestor (black) of the target (red).  (b)
  With the same target in an undirected network, the new node attaches
  to any one of the red neighboring nodes.}
\label{fig:redir-iso}
\end{figure}

The growth rule for isotropic redirection is nearly the same as that
given in Sec.~\ref{sec:redir} for directed networks, but with a small
but profound difference~\cite{krapivsky2017emergent,levens2022friend}
that is embodied by the following growth rule:
\begin{enumerate}
\itemsep -0.5ex
\item Pick a pre-existing node {\tt x} from the network uniformly at random.
\item With probability $1-r$, the new node {\tt n} attaches to ${\tt x}$.
\item Otherwise, with probability $r$, the new node {\tt n} attaches to any neighbor of
  ${\tt x}$, chosen uniformly at random.
\end{enumerate}
Repeat these steps a until a network of a desired size is
generated. The growth rules for directed and undirected redirection
are illustrated in Fig.~\ref{fig:redir-iso}.

\begin{figure}[ht]
\centerline{\includegraphics[width=0.45\textwidth]{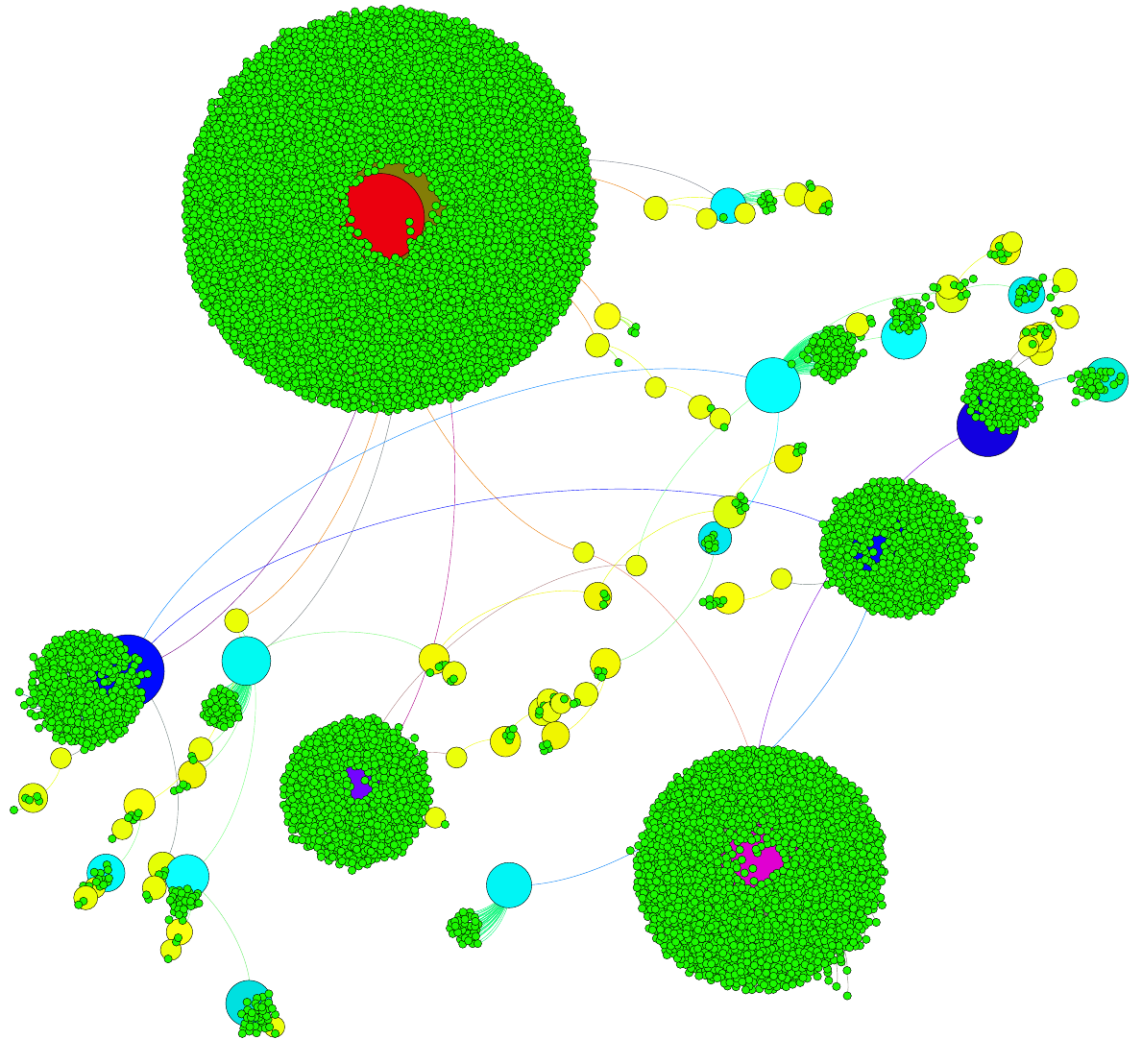}\qquad\includegraphics[width=0.45\textwidth]{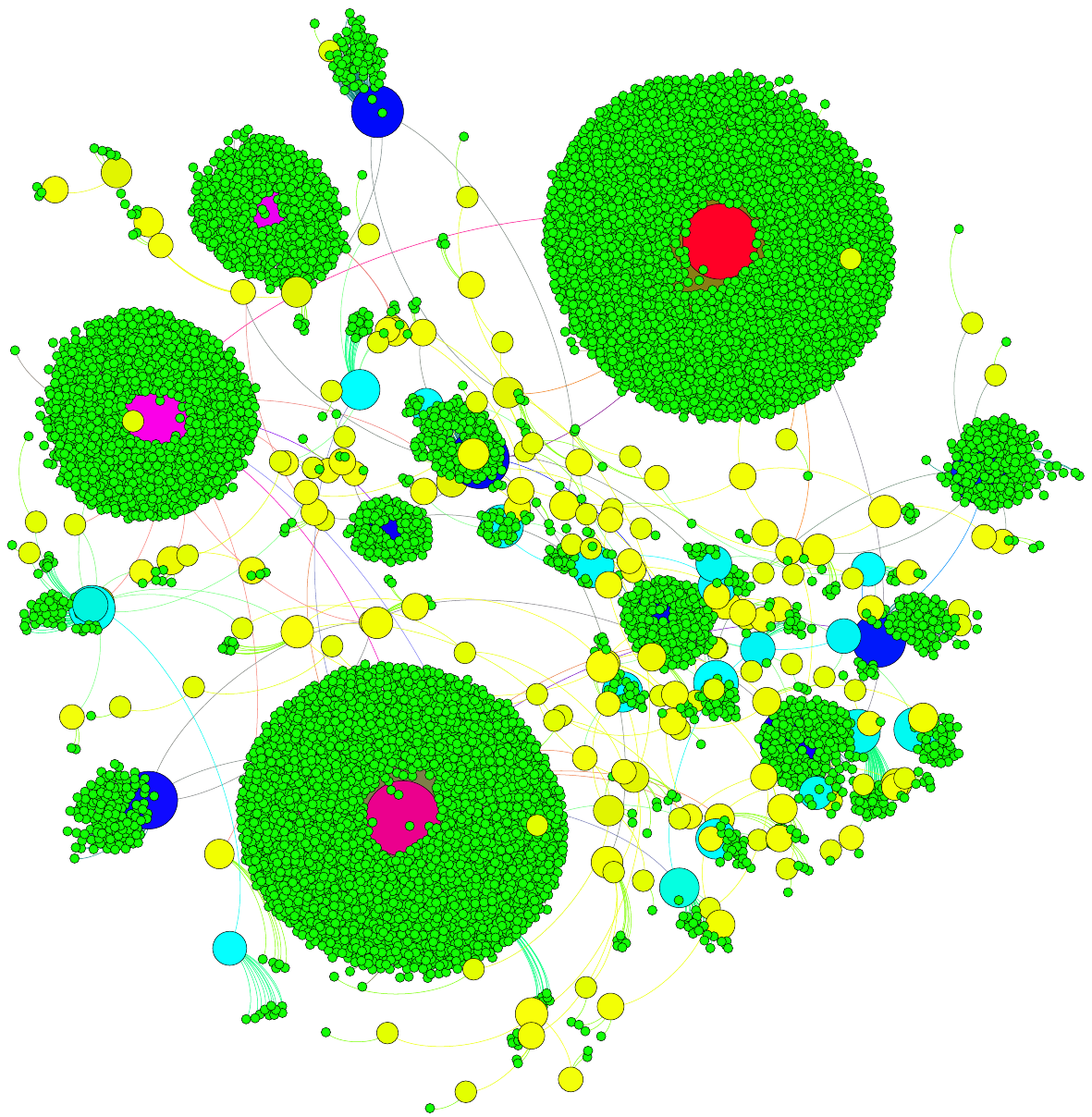}}
\caption{\small Examples of tree networks of $10^4$ nodes that are
  grown by complete redirection.  Green: nodes of degree $k=1$
  (leaves); yellow, $2\!\leq\! k\!\leq\! 10$; cyan,
  $11\!\leq\! k\!\leq\! 99$; blue $100\!\leq\! k\!\leq\! 500$; violet
  $\to$ red, $k\!>\!501$.  The node radius also indicates its
  degree. }
\label{fig:example}
\end{figure}

We focus on the limit of $r=1$, which we term complete redirection,
because this limiting case leads to the most striking phenomenology.
Simulation data also suggest that it is only the special case of $r=1$
that gives rise to emergent modularity.  The network realizations
shown in Fig.~\ref{fig:example} for $r=1$ are highly modular and each
consists of a number of well-resolved modules.  Each module contains a
central macrohub whose degree is a finite fraction of the total number
of nodes $N$; thus each macrohub is connected to a large number of
leaves (nodes of degree 1).  Typical networks consist almost entirely
of leaves as $N\to \infty$; that is, the number of leaves satisfies
$N_1/N\to 1$ as $N\to\infty$.  Nodes with degrees $k\geq 2$ constitute
what we term the network ``core''.  This core comprises an
infinitesimal fraction of the network, viz., the number of core nodes
$\mathcal{C}= \sum_{k\ge 2}N_k$ grows as $N^{\nu-1}$, with
$\nu\approx 1.567$, as determined by numerical
simulations~\cite{krapivsky2017emergent}.

The degree distribution for complete redirection has an algebraic tail
$N_k\propto k^{-\nu}$ with $\nu\approx 1.567$.  As discussed in the
introduction, a degree distribution with $\nu<2$ cannot occur in
sparse networks, which exhibit standard extensive $N_k\propto N$
scaling.  However, a degree distribution with such a fat tail can
arise if the amplitude grows sub-extensively with network size, that
is, $N_k\sim N^{\nu-1}/k^\nu$.  Thus the number of nodes of any
fixed degree $k$ with $k\geq 2$ grows sublinearly in $N$, with
$N_k\sim N^{\nu-1}$.

Several features of networks grown by complete redirection can be
understood analytically \cite{krapivsky2017emergent}, while others,
such as the exponent $\nu$ currently appear to be beyond the reach of
available techniques.  The difficulty in making theoretical progress
is that the change in the degree of a specific node depends on the
degrees of all its neighbors.  This inherent non-locality in the
growth rule means that it is not possible to write a master equation
for the degree distribution alone.  Instead, the equation for the
degree distribution must involve degree correlation functions between
neighboring nodes, and this quantity, in turn, involves higher-order
correlation functions.

\section{Concluding Remarks}

Preferential attachment networks have been the focus of intense
investigation for the past two decades.  Part of the reason for this
explosion of interest stemmed from the confluence of theoretical
insights that were inspired by the existence of new datasets about
networked systems. The network paradigm is alluring and a large number
of seemingly unrelated many-body systems are now studied within the
context of complex networks.

While the field has advanced significantly, some basic facts about the
simplest network models seem under-appreciated.  One is that the degree
distribution of linear preferential attachment networks sensitively
depends on microscopic details of the network growth mechanism.  While
the earliest theoretical studies of linear preferential attachment
networks found a degree distribution exponent of $\nu=3$, any
exponent value with $\nu>2$ can be achieved by linear preferential
attachment.  This non-universality is surprising because the standard
lore from statistical physics suggests that exponent values should be
universal and independent of the details of the network growth
process.

Another important facet of complex networks that has yet to be fully
exploited is that they can be generated by simple redirection
algorithms.  When a new node joins the network, it either attaches
with a given probability to a pre-existing node that is chosen
uniformly at random, or it attaches to the ancestor this target node
with the complementary probability.  This algorithm is simple to
implement and efficient because it generates networks of $N$ nodes in
an algorithmic time that is also of the order of $N$.  We showed how to
generate networks that are equivalent to sublinear preferential
attachment and to shifted linear preferential attachment by suitable
redirection rules.  Our redirection perspective provides crucial
insights that relate the random recursive tree to preferential
attachment networks.

We also briefly discussed undirected networks that are grown by
complete redirection. The resulting networks have a highly modular
structure (Fig.~\ref{fig:example}): the number of core (nodes of
degree $\geq 2$) scales sublinearly with the total number of nodes, as
$N^{\nu-1}$, with $\nu\approx 1.567$.  A natural question here is: why
does this redirection mechanism lead to such singular networks?  We
really don't know.  The master equation approach, which works so well
for directed networks, is inadequate to describe the structure this
class of networks.  This inadequacy stems from the effective
non-locality in the growth mechanism, and different approaches seem to
be needed to truly understand the behavior of this network. An even
deeper reason is the lack of self averaging: the random quantities
$N_k$ for any $k>1$ exhibit huge fluctuations from realization to
realization.  Therefore averages $\langle N_k\rangle$ incompletely
characterize each $N_k$, and, by construction, the master equation
approach only gives average quantities.

The oldest and perhaps still most famous complex network is the
evolving random graph or Erd\H{o}s-R\'{e}nyi (ER) random
graph~\cite{ER60}. The same model appeared earlier in the work of
Flory and Stockmayer~\cite{Flory39,Flory41,Stockmayer43}; this model
turns out to be equivalent to aggregation with the product
kernel~\cite{KRB}. The percolation transition manifested by the
emergence of the giant component~\cite{Janson93} in evolving random
graphs is equivalent to gelation in aggregation \cite{KRB}.  The ER
random graph initially consists of $N$ disjoint nodes, and it evolves
by drawing randomly chosen pairs of nodes and connecting them. Thus
only the number of links increases.  Combining the ER graph with
preferential attachment, one may postulate that nodes of degree $i$
and $j$ connect with probability proportional to
$(i+\lambda)(j+\lambda)$. This evolving graph undergoes a percolation
transition and later a condensation transition when the entire system
condenses into a single component~\cite{BK12,10.1214/20-AAP1610}.

Closer to our modeling is a network that grows via two distinct
mechanisms: (i) a new node is added with probability $p$, and (ii) a
new link between existing nodes is created with probability $1-p$.
Both of these steps can incorporate redirection in a natural way.
Earlier work on similar models~\cite{KRR} was focused on network
characteristics, such as the degree distribution. The distribution of
components remains mostly unexplored and it would be interesting to
analyze percolation and condensation transitions for this type of
network. There are indications~\cite{KRR} that the percolation
transition could be different from the standard Curie-type transition
appearing in the ER graphs~\cite{Janson93}, viz., a
Berezinskii-Kosterlitz-Thouless infinite-order transition
\cite{BKT-Strogatz,KR01-Rodgers,BKT-Sergey,KR02-protein} that often
appears in growing networks.

\bigskip\noindent This research was partially supported by various NSF
awards over the past two decades, and most recently by NSF grant
DMR-1910736.



%
\end{document}